\documentclass[acmtog, nonacm]{acmart}

\AtBeginDocument{%
  \providecommand\BibTeX{{%
    \normalfont B\kern-0.5em{\scshape i\kern-0.25em b}\kern-0.8em\TeX}}}

\usepackage{pifont}

\usepackage{subcaption}
\usepackage{colortbl}

\usepackage{listings}
\usepackage{xcolor}

\definecolor{codegreen}{rgb}{0,0.6,0}
\definecolor{codegray}{rgb}{0.5,0.5,0.5}
\definecolor{codepurple}{rgb}{0.58,0,0.82}
\definecolor{backcolour}{rgb}{0.99,0.99,0.99}

\lstdefinestyle{mystyle}{
  backgroundcolor=\color{backcolour},   
  commentstyle=\color{codegreen},
  keywordstyle=\color{magenta},
  numberstyle=\tiny\color{codegray},
  stringstyle=\color{codepurple},
  basicstyle=\ttfamily\footnotesize,
  breakatwhitespace=false,         
  breaklines=true,                 
  captionpos=b,                    
  keepspaces=true,                 
  numbers=left,                    
  numbersep=5pt,                  
  showspaces=false,                
  showstringspaces=false,
  showtabs=false,                  
  tabsize=2, 
  moredelim=**[is][\color{red}]{@}{@},
}

\lstdefinelanguage{yaml}{
     basicstyle=\footnotesize\ttfamily\color{blue},
     rulecolor=\color{black},
     string=[s]{'}{'},
     stringstyle=\color{blue},
     comment=[l]{:},
     commentstyle=\color{black},
     morecomment=[l]{-}
 }

\newcommand{\cmark}{\ding{51}}%
\newcommand{\xmark}{\ding{55}}

\definecolor{dgreen}{RGB}{61,152,56}
\definecolor{redcode}{RGB}{170,73,38}

\newcommand\new[1]{\textcolor{blue}{\textbf{#1}}}
\newcommand\old[1]{\textcolor{red}{\textit{#1}}}

\newcommand{\fossil}{\textsf{Fossil }}
\newcommand{\cX}{\mathcal{X}}
\newcommand{\cI}{\mathcal{X}_I}

\newcommand{\cG}{\mathcal{X}_G}
\newcommand{\cS}{\mathcal{X}_S}
\newcommand{\cU}{\mathcal{X}_U}
\newcommand{\cF}{\mathcal{X}_F}
\newcommand{\real}{\mathbb{R}}
\newcommand{\extreal}{\overline{\mathbb{R}}}

\begin{document}

\title{\fossil 2.0: Formal Certificate Synthesis for the Verification  and Control of Dynamical Models}

\author{Alec Edwards}
\affiliation{%
  \institution{University of Oxford}
  \streetaddress{Parks Road}
  \city{Oxford}
  \country{UK}}
\email{alec.edwards@cs.ox.ac.uk}
\orcid{0000-0001-9174-9962}

\author{Andrea Peruffo}
\affiliation{%
  \institution{TU Delft}
  \city{Delft}
  \country{the Netherlands}
}
\email{a.peruffo@tudelft.nl}
\orcid{0000-0002-7767-2935}

\author{Alessandro Abate}
\affiliation{%
 \institution{University of Oxford}
 \streetaddress{Parks Road}
 \city{Oxford}
 \country{UK}}
 \orcid{0000-0002-5627-9093}

\renewcommand{\shortauthors}{Edwards, et al.}

\begin{abstract}
    This paper presents \fossil 2.0, a new major release of a software tool for the synthesis
    of certificates (e.g., Lyapunov and barrier functions) for dynamical systems modelled as ordinary differential and difference equations.  \fossil 2.0 is much improved from its original release, including new interfaces, a significantly expanded certificate portfolio, controller synthesis and enhanced extensibility. We present these new features as part of this tool paper.
    \fossil implements a counterexample-guided inductive synthesis (CEGIS) loop ensuring the soundness of the method.
    Our tool uses neural networks as templates to generate candidate functions, which are then formally proven by an satisfiability modulo theories solver acting as an assertion verifier. 
    Improvements with respect to the first release include 
    a wider range of certificates, 
    synthesis of control laws, 
    and 
    support for discrete-time models.  
\end{abstract}

\begin{CCSXML}
<ccs2012>
   <concept>
       <concept_id>10010147.10010257.10010293.10010294</concept_id>
       <concept_desc>Computing methodologies~Neural networks</concept_desc>
       <concept_significance>500</concept_significance>
       </concept>
   <concept>
       <concept_id>10002944.10011123.10011676</concept_id>
       <concept_desc>General and reference~Verification</concept_desc>
       <concept_significance>500</concept_significance>
       </concept>
   <concept>
       <concept_id>10010520.10010553</concept_id>
       <concept_desc>Computer systems organization~Embedded and cyber-physical systems</concept_desc>
       <concept_significance>500</concept_significance>
       </concept>
 </ccs2012>
\end{CCSXML}

\ccsdesc[500]{Computing methodologies~Neural networks}
\ccsdesc[500]{General and reference~Verification}
\ccsdesc[500]{Computer systems organization~Embedded and cyber-physical systems}

\keywords{Lyapunov-like functions, CEGIS, SAT-modulo theories, Computer-aided control design, 
Neural networks}

\maketitle

\section{Introduction}
\label{sec:intro}

This paper describes a major new release, version 2.0, of the tool \textsf{Fossil}, a software package for the sound synthesis of certificates aimed at the verification and control of dynamical models.
Moving much beyond the earlier release, which solely encompassed the synthesis of Lyapunov and barrier functions for continuous-time dynamical models, 
\fossil 2.0 greatly extends the set of considered specifications, which encode requirements or properties for the models under study: in particular dealing with the computation of certificates for ROA (proving a set is (inside) a region of attraction), SWA (stable while avoid), RWA (reach while avoid), RSWA (reach and stay, while avoid), RAR (reach-avoid and remain). 
A second major extension of the tool is the synthesis of (neural) control laws. This is done concurrently with certificate synthesis, allowing for \fossil 2.0 to learn control laws which guide models to satisfy a specification at the same time as synthesising certificates which prove the resulting closed loop model satisfies the specification. These control laws take the form of neural networks; whether the resulting closed loop neural ODE satisfies the desired specification is the purpose of the concurrent certificate synthesis.
We do not refer to these as control certificates, which are instead certificates which prove that for each state there exists a control input that will allow the specification to hold.

The codebase has been significantly refactored and rewritten, allowing the implementation of several innovations, besides of course handling the much richer set of specifications. 
Its novel structure improves the tool's extensibility and  
it bridges the gap between formal specifications and controller synthesis.  
Further, 
\textsf{Fossil 2.0} enjoys a brand-new command-line interface, allowing a more direct way to synthesise a desired certificate. 
Finally, the verification engine has been augmented to include a new Satisfiability Modulo Theories (SMT) solver, CVC5.

Synthesis with neural architectures is a growing trend in the control literature, 
with a wide range of applications, e.g., fault tolerant control \cite{grande2023passsive}, robotics and multi-agent systems \cite{dawson2022safe}, safety for stochastic systems \cite{mathiesen2023safety, Žikelić_Lechner_Henzinger_Chatterjee_2023, Lechner_Žikelić_Chatterjee_Henzinger_2022, chatterjee2023LearnerverifierFrameworkNeural}. 
In the context of certificates synthesis, 
most of the existing work has focused on Lyapunov and barrier functions, 
however it is evident that complex applications require richer specifications, e.g., reaching a target region while avoiding an undesirable (or unsafe) portion of the state space: this requirement is embodied by the RWA certificate in our tool, which goes much beyond this single objective and indeed aims at the composition of certificates for ever richer requirements. 
Early works on sound Lyapunov and barrier function synthesis
can be found in \cite{ahmed2018AutomatedSoundSynthesis, chang2020NeuralLyapunovControl, abate2020AutomatedFormalSynthesis, samanipour2023StabilityAnalysisController, grande2023augmented, grande2023passsive,peruffo2021AutomatedFormalSynthesis, zhao2020SynthesizingBarrierCertificates, ratschan2018SimulationBasedComputation}.
More complex properties, such as `reach while stay', are discussed in \cite{verdier2020FormalControllerSynthesis, verdier2020FormalSynthesisAnalytic,ravanbakhsh2015CounterexampleGuidedSynthesisReachWS}, 
while a recent survey on neural certificates is presented in \cite{dawson2022safe}. 

Fossil 1.0 \cite{abate2021FOSSILSoftwareTool} is a tool 
based upon earlier works on neural template synthesis for Lyapunov \cite{ahmed2018AutomatedSoundSynthesis, abate2020AutomatedFormalSynthesis} and barrier functions \cite{peruffo2021AutomatedFormalSynthesis}. Within this narrower focus, these works benchmark Fossil against alternative synthesis techniques, such as SOStools \cite{papachristodoulou2013SOSTOOLSVersion00} and neural Lyapunov control (NLC) \cite{chang2020NeuralLyapunovControl}, proving that  it outperforms them in terms of computational time, robustness, whilst supporting a larger set of characteristics (cf. Table \ref{tab:meth-characteristics}). 
Recently, a general verification framework for dynamical models via inductive synthesis of certificate via counter-examples (CEGIS) has been introduced:   \cite{edwards2023GeneralVerificationFramework} presents a 
theoretical framework 
for controller and certificate synthesis, for a broad range of properties (requirements). 
The present tool paper is built upon the methodology outlined in \cite{edwards2023GeneralVerificationFramework} by introducing a new, user-friendly tool.  
We detail the novel contributions of this tool next.

\begin{table*}[htbp]
    \centering
    
\begin{tabular}{l|lllllll|lllllll}
    \toprule
    & \multicolumn{6}{c}{\textbf{Characteristics}} & \multicolumn{7}{c}{\textbf{Properties}} \\ \midrule
   & \rotatebox{90}{Polynomial} & \rotatebox{90}{Non-polynomial} & \rotatebox{90}{User Interface} & \rotatebox{90}{Nonlinear Ctrl} & \rotatebox{90}{Sound} & \rotatebox{90}{Non-convex} & \rotatebox{90}{Discrete time}
   & \rotatebox{90}{Stability} & \rotatebox{90}{ROA} & \rotatebox{90}{Safety} & \rotatebox{90}{SWA} & \rotatebox{90}{RWA} & \rotatebox{90}{RSWA} & \rotatebox{90}{RAR} 
    \\ \midrule
    {Fossil 2.0}
    & \cmark & \cmark & \cmark & \cmark & \cmark & \cmark & \cmark 
    & \cmark & \cmark  & \cmark & \cmark & \cmark & \cmark & \cmark    
    \\
    {Fossil 1.0 \cite{abate2021FOSSILSoftwareTool}}
    & \cmark & \cmark & \cmark & \xmark & \cmark & \cmark & \xmark
    & \cmark & \xmark & \cmark & \xmark & \xmark & \xmark & \xmark            
        \\
    {F4CS \cite{verdier2020FormalControllerSynthesis, verdier2020FormalSynthesisAnalytic}} 
    & \cmark & \xmark & \xmark & \cmark & \cmark & \cmark & \xmark
    & \cmark & \xmark & \cmark & \xmark & \cmark & \cmark & \xmark 
        \\
    NLC \cite{chang2020NeuralLyapunovControl} 
    & \cmark & \cmark & \xmark & \cmark & \cmark & \cmark & \xmark
    & \cmark & \xmark & \xmark & \xmark & \xmark & \xmark & \xmark  
        \\
    SOStools \cite{papachristodoulou2013SOSTOOLSVersion00} 
    & \cmark & \xmark & \cmark & \cmark & \xmark & \xmark & \cmark  
    & \cmark & \xmark & \cmark & \xmark & \xmark & \xmark & \xmark  
        \\
    \bottomrule 
\end{tabular}

    \caption{Characteristics of certificate synthesis approaches works across literature. Included are works with an attached codebase, regardless of the existence of a repeatability package or the maintenance of the code itself.
    On the right-hand side, properties that can be verified by the corresponding tool. 
    } 
    \label{tab:meth-characteristics}
\end{table*}

\begin{table}[htbp]
    \centering
    \begin{tabular}{l|l}
    \toprule
     Feature & Details  \\
     \midrule

     Interface & \old{Jupyter Interface}, \new{Command Line},\\
     &  \new{Python Interface} \\
     Properties & Stability, Safety, \new{SWA}, \new{RWA}, \new{RSWA}, \new{ROA}, \new{RAR}, \\
     Models & Continuous Time, \new{Discrete Time} \\
     Verifiers & Z3, dReal, \new{CVC5} \\
     Domains & Spheres, Boxes, \new{Open spheres}, \new{Open boxes}, \\
     & \new{Ellipsoids}, Custom sets \\
     Misc. & \new{Control Synthesis}, \\
     & \new{Certificate Extensibility}, \new{Verifier-only}, \\& \new{Learner-only}
    
\end{tabular}
    \caption{Comparison of features between the two releases of Fossil. Italic red text denotes Fossil 1.0 only, and bold blue text denotes Fossil 2.0 only.}
    \label{tab:tool-features}
\end{table}

\subsection{Overview of Functionality}
\label{subsec:function-overview}

\begin{figure*}[htbp]
    \centering
    \resizebox{0.9\textwidth}{!}
    {
    \includegraphics{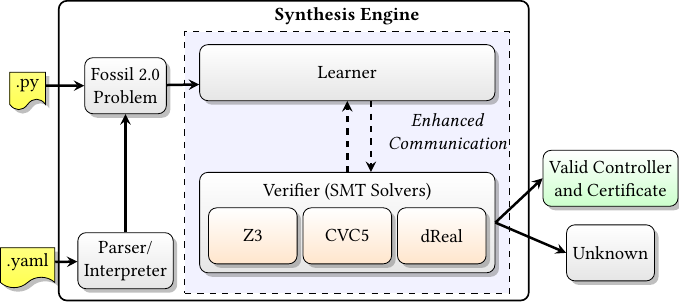}
    }
    \caption{General architecture of Fossil 2.0.}
    \label{fig:fossil-arch}
\end{figure*}

We begin with a brief overview of the current functionality of Fossil, and present a high-level overview of its architecture in Figure \ref{fig:fossil-arch}. 
We also summarise these features, and how they compare to the features of the original Fossil 1.0 release, in Table \ref{tab:tool-features}. 

\begin{itemize}
    \item  a tool for robust formal synthesis of certificates of elaborate specifications categorised as  \textbf{reachability}, \textbf{avoidance}, \textbf{remain}, alongside the concurrent synthesis of neural-network \textbf{controllers} for these specifications (see Table~\ref{tab:meth-characteristics} for a summary) ; 
    \item an easy-to-use \textbf{command-line interface} and a Python-based interface for usage of Fossil 2.0
    \item verification of more properties, and additional model types (\textbf{discrete-time}) for Lyapunov and barrier functions relative to \fossil 1.0;
    \item a correspondingly broader \textbf{benchmark suite} of dynamical models and associated properties over \fossil 1.0. 
\end{itemize}

Table~\ref{tab:meth-characteristics} summarises the properties under study, and the features of the methodology used in \textsf{Fossil 2.0} relative to works in the literature. We only include approaches with an existing code-base or software, though this does not imply the existence of a repeatability package. 
In particular, \textsf{Fossil 2.0}'s methodology enables polynomial and non-polynomial certificates, with linear and non-linear control design, over convex and non-convex sets. Finally, let us emphasise that, in view of the SMT verification engine underpinning it, our tool is \emph{sound}, namely the result of the synthesis is formally valid over (a dense domain within) $\mathbb{R}^n$ - this is unlike many alternatives in the literature.

\section{Tool Scope and Breadth}
\label{sec:tool-scope-and-prop}

\subsection{Scope}
\label{subsec:scope}

We formulate several fundamental problems in the analysis and verification of dynamical systems modelled as ordinary differential equations (ODEs) or as ordinary difference equations in terms of \emph{reaching} (either in finite-time, or asymptotically) a desired set, of \emph{avoiding} an unsafe/undesired set, and of \emph{remaining} within a final set. 
These properties (requirements) are core to control theory, involve a multitude of practical applications, from robotic tasks to self-driving vehicles \cite{dawson2022safe}, and can be formally encoded as temporal specifications.  
We are thus interested in sufficient conditions certifying relevant specifications of dynamical models: 
to this end, we consider the synthesis of certificates functions, whose most notable examples are Lyapunov and barrier functions. 
In practical terms, the existence (and thus practically, the construction) of (any) such function(s) ensures the satisfaction of the specification for the given dynamical model. Ultimately, we may need to additionally synthesise a control law to fulfil the given specification. 

The reachability property requires trajectories to arrive at a given target set, either in finite time (the most canonical definition of reachability, see e.g. \cite{aastrom2021feedback, baier2008principles})
or asymptotically (which is classically shown via Lyapunov techniques). 
Avoidance is typically shown via barrier functions, whereby all trajectories starting from a given initial set should never reach an unsafe/undesired set. 
The combination of these two properties give rise to the stay while avoid (SWA) and reach while avoid (RWA) certificates. 
After having reached a target set, trajectories might be required to dwell within a final, possibly different, set: this is denoted as the remain property: certificates such as RSWA and RAR derive from the combination of these three properties.

\subsection{Properties}
\label{subsec:properties}

\begin{figure*}[htbp]
    \centering
    \resizebox{\textwidth}{!}{
         \subfloat[][Stability]{\includegraphics{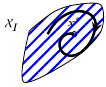} \label{fig:stab}}
         \subfloat[][ROA]{\includegraphics{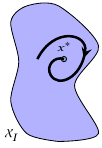} \label{fig:roa}}
         \subfloat[][Safety]{\includegraphics{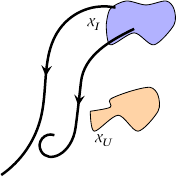} \label{fig:safe}}
         \subfloat[][SWA]{\includegraphics{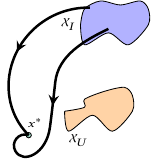} \label{fig:swa}}
         \subfloat[][RWA]{\includegraphics{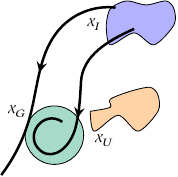} \label{fig:rwa}}
         \subfloat[][RSWA]{\includegraphics{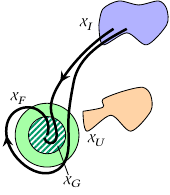} \label{fig:rswa}}
         \subfloat[][RAR]{\includegraphics{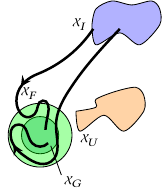} \label{fig:rar}}
         }
     \caption{Pictorial depiction of relevant properties verifiable by \fossil 2.0. Here, $\cI$ is the initial set, $\cU$ the unsafe set ($\cS$ is its safe complement), $\cG$ the goal/target set, $\cF$ the final set. (The entire state space is $\cX$.) A dashed background denotes that the corresponding set's existence is implied by the corresponding certificate, but that it is not explicitly defined in the property. }
     \label{fig:properties-example}
\end{figure*}

Let us now outline the rich portfolio of trajectory-based properties that \fossil 2.0 supports. We note that \fossil 1.0 is limited to Lyapunov functions and barrier certificates, and therefore supports only two of the properties described in this section, namely \emph{stability} and \emph{safety}. Note that we only present the properties verified by our tool: the actual certificates that are synthesised to prove these properties are not discussed for brevity, and the interested reader is directed to the technical details in \cite{edwards2023GeneralVerificationFramework}. The properties presented are summarised and illustrated in Figure \ref{fig:properties-example}.

We 
define properties over trajectories of dynamical models, which suits both continuous- and discrete-time dynamics. 
In the following, we consider models described by 
\begin{equation}
    \label{eq:con-model}
    \dot{\xi}(t) = 
    f(\xi(t), u(t)), 
    \quad 
    \xi(t_0) = x_0 \in \cI \subseteq \cX, 
\end{equation}
where $x \in \cX \subseteq \mathbb{R}^n$ is the state of the system, with initial set $\cI$, 
$u \in \cU \subseteq \mathbb{R}^m$ is the control input, 
$f : \cX \times \mathcal{U} \rightarrow \mathbb{R}^n$ is a Lipschitz-continuous vector field describing the model dynamics. 
Further, we denote the unsafe set  as $\cX_U$, representing a region of the state space that the system's trajectories should avoid; 
$\cX_G$ represents a goal set, indicating the region that the system's trajectories should enter;
and $\cX_F$ represents a final set, indicating a set where the system's trajectories should remain for all times after arriving at the goal set. 
We also assume the following relations among sets hold: 
$\cU \cap \cF = \emptyset$ and $\cG \subset \cF$.
Given a set $S$ in a domain $\cX$, we denote by ${S}^\complement$ its complement, i.e. $\cX \setminus S$, and by $int(S)$ its interior, namely the set without its border, i.e. $int(S) = S \setminus \partial S$. 
Given a control law $\overline{u}$, we assume $f(\xi, \overline{u})$ has an equilibrium point $x^*$. %
The special case of autonomous models is obtained by considering trivial control inputs and sets. For simplicity, we present properties for autonomous models, as the extension of their  definition to control models is straightforward - but much less so is the actual combined synthesis of controllers and certificates, as discussed below.

\paragraph{Lyapunov Stability}

Stability is  arguably the most widely studied property of dynamical models, and is often characterised in a Lyapunov (i.e. asymptotic) sense \cite{sastry1999NonlinearSystems}, namely in terms of the distance of a trajectory from an  equilibrium point of the model. Qualitatively, it requires that any trajectory starting within an unspecified initial set $\cI$ (containing $x^*$) eventually reaches $x^*$. 
We formally express this as 
\begin{equation}
\label{eq:stab-spec}
        \exists \cI \subseteq \cX: \forall \xi(t_0) \in \cI, \exists T \in \extreal, \forall \tau \geq T, \ \xi(\tau) \in \{x^*\},
\end{equation}
where $\cI$ (notice the existential quantifier) has non empty interior. 
For the sake of clarity, the conditions on a Lyapunov certificate for stability are displayed below, in Equations \eqref{eq:lyap}.

\paragraph{Region of Attraction (ROA)}

A Lyapunov function proves the \emph{existence} of some region \emph{within} $\cX$, within which initialised trajectories will converge asymptotically towards the origin. This is known as a \emph{region of attraction} (ROA). 
Oftentimes we are interested in proving that all trajectories initialised within a \emph{given set} $\cI$ converge to the equilibrium. 
This is a different problem to that of stability, and therefore requires a different specification (and corresponding certificate). 
We thus subtly modify \eqref{eq:stab-spec} to require a specified or given set of initial states $\cI$, which should be a region of attraction for an equilibrium point, as follows:
\begin{equation}
\label{eq:roa-spec}
        \text{(Given) } \cI, \ 
        \forall \xi(t_0) \in \cI, \exists T \in \extreal, \forall \tau \geq T, \ \xi(\tau) \in \{x^*\}. 
\end{equation}

\paragraph{Barrier Functions for Safety}

Safety properties specify that no trajectory starting from an initial set $\cI$ may enter some unsafe set $\cU$; for continuous-time and discrete-time models, safety over an unbounded time horizon can be proved via barrier certificates \cite{prajna2004StochasticSafetyVerification, prajna2006BarrierCertificatesNonlinear}. Here, we consider an unbounded-time safety specification
Formally, the trajectories fulfil the following property
\begin{equation}
\label{eq:safety-spec}
    \forall \xi(t_0) \in \cI, \forall t  \in \extreal, t \geq t_0, \xi(t) \in {\cU}^\complement.
\end{equation}

\paragraph{Stable While Avoid}

The conjunction of the conditions of ROA and safety is also a salient property. This specifies that 
all trajectories starting in some initial set converge towards an equilibrium point while also avoiding a given unsafe set. 
Certifying this is equivalent to concurrently certifying both stability \eqref{eq:roa-spec} and safety \eqref{eq:safety-spec} hold; the behaviour can be expressed as 
\begin{multline}
\label{eq:stab-safe-spec}
    \forall \xi(t_0) \in \cI, \exists T \in \extreal, \forall t \in [t_0, T), \xi(t) \in {\cU}^\complement \\
    \wedge \forall \tau \geq T,
    \xi(\tau) \in \{x^*\}.
\end{multline}

\paragraph{Reach While Avoid}

A reach while avoid (RWA) property involves the avoidance of an unsafe set whilst guaranteeing \emph{finite-time} reachability to some goal region.
Let us 
define an unsafe set $\cU = \cX \setminus \cS$, where $\cS$ is a compact safe set, a compact initial set $\cI \subset int(\cS)$, and a compact goal set $\cG \subset int(\cS)$ with non-empty interior.
In formal terms, 
\begin{equation}
\label{eq:rwa-spec}
    \forall \xi(t_0) \in \cI, \exists T \in \real, \forall t \in [t_0, T] :
    \
    \xi(t) \in {\cU}^\complement \wedge \xi(T) \in \cG.
\end{equation}
Note that a RWA property does \emph{not} require that trajectories will remain within the goal set, or that trajectories shall avoid the unsafe set for all (unbounded) time. 
It is possible for trajectories to leave the goal set after entering it, and thereafter possibly enter the unsafe set $\cU$. This scenario is addressed by the next two properties.

\paragraph{Reach-and-Stay While Avoid}

A reach-and-stay while avoid (RSWA) property modifies a  RWA property by ensuring that trajectories will remain within some final (or terminal) set for all time, and is described as follows:
This property does not require that trajectories reach the final set and immediately remain within it. In fact, trajectories may enter and leave the final set as long as at some point they enter and never leave again.
However, in finite time trajectories must reach some subset of the final set a goal set, after which they must remain within the final set for all time. This goal set is not explicitly specified, but it must exist.
Formally, trajectories should satisfy the property
\begin{multline}
   \label{eq:rswa-spec} 
    \exists \cG: \forall \xi(t_0) \in \cI, \exists T \in \real, \forall t \in [t_0,T], \xi(t) \in {\cU}^\complement \\
    \wedge \xi(T) \in \cG
    \wedge \forall \tau \geq T: \xi(\tau)\in \cF.
\end{multline}

\paragraph{Reach, Avoid and Remain}
A Reach Avoid Remain (RAR) specification entails a RSWA property, but for given goal and final sets. In other words, we remove the existential quantification over the goal set and seek to specify this set explicitly.
We express this formally, as follows: 
\begin{multline}
\label{eq:rar-spec}
    \forall \xi(t_0) \in \cI, 
    \exists T \in \real, \forall t \in [t_0, T]: 
    \\
    \xi(t) \in {\cU}^\complement \wedge \xi(T) \in \cG \wedge \forall \tau \geq T: \xi(\tau) \in \mathcal{X}_F. 
\end{multline}

\subsection{A Note on Control Models}
\label{subsec:control}

The canonical design procedure \cite{sontag2013mathematical} of, e.g., control Lyapunov function, 
in the first instance defines a Lyapunov function, and 
secondly chooses the control inputs allowing for the (stability) property to hold. 
A control Lyapunov function therefore proves that there always exists a suitable control input such that the Lyapunov conditions hold.
This approach entails that, after a candidate control-certificate is provided, control actions can be determined, 
e.g. by solving an optimisation program over the input space and the candidate certificate. 
This is \emph{not} the approach taken by this tool. Instead, Fossil 2.0 synthesises a feedback control law for the model described by  \eqref{eq:con-model}, and ``applies'' this state feedback to obtain a closed-loop model, for which we synthesise a certificate. Whilst we synthesise the control law \emph{concurrently} with the certificate, we do not refer to these as ``control certificates'', as in literature. Hence, we verify properties for control models using both a controller and a certificate, and in particular do not delegate the controller synthesis a-posteriori. 
 
In this release of \fossil, we devise a dedicated control loss function to penalise trajectories that stray away from the origin (we assume that the goal, target sets contain the origin),  
inspired by the \emph{cosine similarity} of the two vectors $d$ and $f(d)$, and rewards vectors for pointing in opposite directions \cite{edwards2023GeneralVerificationFramework}. 
This loss function only concerns the dynamics $f(\xi, u)$ and disregards the candidate certificate, as it specifically focuses on the parameters in the feedback law to learn a desirable $f(\xi, u)$. 
We could also add other desirable features to this control loss function: for instance, a quadratic term to penalise large values of the control input $u(t)$, mimicking an LQR. Naturally, these represent soft-constraints, as the control law might indeed not follow exactly the user-defined constraints.

\subsection{Inductive Certificate Synthesis with Counter-Examples}
\label{subsec:cegis}

Fossil 2.0 leverages an automated and formal approach for the construction of certificates that are expressed as feed-forward neural networks.
The procedure is built on CEGIS,  an automated and sound procedure for solving second-order logic synthesis problems, which comprises two interacting parts, as outlined in Fig.~\ref{fig:fossil-arch}. 
The first component is a \emph{learner} based upon neural network templates, which acts in a numerical environment,
trains a candidate to satisfy the conditions over a finite set $D$ of samples.  The second component, which works in a symbolic environment, is a \emph{verifier} that either confirms or falsifies whether the conditions are satisfied over the whole dense domain $\cX$. If the verifier falsifies the candidate, %
one or more counter-examples are added to the samples set and the network is retrained. 
This loop repeats until the verification proves that no counter-examples exist or until a timeout is reached.

The performance of the CEGIS algorithm in practice hinges on the effective exchange of information between the learner and the verifier. The CEGIS architecture within Fossil is tailored to provide enhanced communication between numerical and symbolic domains, which is achieved through dedicated subroutines. For more details regarding the augmented CEGIS loop, the interested reader is referred to \cite{abate2021FOSSILSoftwareTool}.

\section{Tool Use -- Operating Instructions}
\label{sec:use}

\subsection{Command Line Interface}
\label{subsec:use-cli}

\fossil 2.0 is endowed with an \emph{easy-to-use} interface based on command line, which leverages YAML configuration files to define the required parameters for the program. We guide through its use with two test benchmarks: an autonomous and a control model.

\subsubsection{Simple Use-case}
Let us consider the following continuous-time dynamical model,
\begin{equation} 
\label{eq:ex-dyn-1}
    \begin{cases}
        \dot{x}_0 = x_1 - x_0^3, \\
        \dot{x}_1 = -x_0,
    \end{cases}
\end{equation}
which has a single equilibrium located at the origin. We can use \fossil to prove whether this equilibrium is (locally) asymptotically stable  by synthesising a Lyapunov function.
To this end, it is sufficient to define a YAML file as follows: 

\begin{lstlisting}[language=yaml, caption={Example YAML configuration file to synthesise a Lyapunov function using Fossil 2.0.}, label={lst:lyap-yaml}]
N_VARS: 2
SYSTEM: [x1 - x0**3, -x0]
CERTIFICATE: Lyapunov
TIME_DOMAIN: CONTINUOUS
DOMAINS:
  XD: Sphere([0,0], 1.0)
N_DATA:
  XD: 1000
N_HIDDEN_NEURONS: [5]
ACTIVATION: [SQUARE]
VERIFIER: Z3
\end{lstlisting}

We specify the system dynamics and certificate type in the corresponding fields.
Since certificates in Fossil are neural networks, we must input their structure as part of the configuration. In this example, we specify a network consisting of a single hidden layer (5 neurons) with quadratic (square) activation functions (resulting in an SOS-like quadratic Lyapunov function). We outline the domain of verification (which implicitly impacts the verified region of attraction) as a hyper-sphere, centred at the origin of radius $1.0$. We then specify that $1000$ data points should be sampled from this domain to train the Lyapunov function. Finally, we tell Fossil to perform the verification step using Z3.

\subsubsection{Controller Synthesis}
Fossil 2.0 is able to synthesise feedback controllers for dynamical models with control input. These controllers are synthesised concurrently with a certificate, and guide the model to satisfy the required conditions. Consider a modified version of the model in \eqref{eq:ex-dyn-1} as:
\begin{equation} 
\label{eq:ex-dyn-ctrl}
    \begin{cases}
        \dot{x}_0 = x_1 - x_0^3, \\
        \dot{x}_1 = u_0,
    \end{cases}
\end{equation}
where $u_0$ represents a control input. We can modify the configuration file in Listing \ref{lst:lyap-yaml} to synthesise a simple linear controller and Lyapunov function for this model, which we show in Listing \ref{lst:ctrl-lyap-yaml}.

\begin{lstlisting}[language=yaml, caption={Example YAML configuration file to synthesise a Lyapunov function and corresponding feeback controller using Fossil 2.0.}, label={lst:ctrl-lyap-yaml}]
N_VARS: 2
SYSTEM: [x1 - x0**3, u0]
CERTIFICATE: Lyapunov
TIME_DOMAIN: CONTINUOUS
DOMAINS:
  XD: Torus([0,0], 1.0, 0.01)
N_DATA:
  XD: 1000
N_HIDDEN_NEURONS: [5, 5]
ACTIVATION: [SIGMOID, SQUARE]
CTRLAYER: [5,1]
CTRLACTIVATION: [LINEAR]
VERIFIER: DREAL
\end{lstlisting}

In this example, we use dReal as a verifier. 
In view of the internal mechanics of dReal ($\epsilon$-satisfiability, cf. \cite{gao2019NumericallyrobustInductiveProof}), we should exclude a small region around the origin from the domain, to avoid pathological problems involving the equilibrium point (which is the origin here). 
This issue is limited to Lyapunov certificate synthesis using dReal. 
We overcome this impediment by employing a spherical domain where a smaller, inner spherical region is removed -- e.g., in two dimensions, this results in an annulus. 
\fossil supports this feature with the domain denoted \texttt{Torus($c$, $r_o$, $r_i$)} --- a slight abuse of nomenclature --- which refers to the hyper sphere centred at $c$ of radius $r_o$ (set)-minus the hyper-sphere centred at $c$ of radius $r_i$. This domain grants the so-called $\epsilon$-stability -- the interested reader may refer to \cite{gao2019NumericallyrobustInductiveProof} for a detailed outlook.

Fossil is able to synthesise complex neural network controllers (e.g., polynomial and non-polynomial designs), though here we declare a simple linear feedback law. Meanwhile, we specify that the certificate should consist of two hidden layers: one of sigmoidal activation functions and one of square activations.

\subsection{Advanced (Python-based) Interface}
\label{subsec:use-python-interface}

Our command line interface is comprehensive, providing users with the ability to synthesise any of Fossil 2.0's seven certificates alongside control laws. 
Fossil may also be interfaced as a Python package, allowing for a more feature-rich experience in terms of functionality and extensibility.

Let us now describe the definition of the synthesis procedure for the model in \eqref{eq:ex-dyn-ctrl} within Python. 
The definition of a model requires a class environment equipped with two methods, as follows.
\begin{lstlisting}[language=Python, caption={Example model definition.}, label=lst:model-def]
import fossil 

class TestModel(fossil.control.ControllableDynamicalModel):
    n_vars = 2  # system variables
    n_u = 1  # control inputs
    
    def f_torch(self, v, u):  # tensor computations
        x0, x1 = v.T
        u0 = u[:,0]
        return [x1 - x0**3, u0]
    
    def f_smt(self, v, u):  # smt computations
        x0, x1 = v
        u0 = u
        return [x1 -x0**3, u0]
\end{lstlisting}
\begin{sloppypar}
Within Fossil 2.0, dynamical models may be declared as objects inheriting from either the \textsf{DynamicalModel} class (for simply autonomous models) and \textsf{ControllableDynamicalModel} (for models with control input to be realised as a state-feedback law).  
The class presents the number of variables and control inputs as \texttt{n\_vars} and \texttt{n\_u}, respectively; autonomous models do not need the instantiation of \texttt{n\_u}. 
The two specular methods define the dynamical model, to be manipulated by PyTorch (\texttt{f\_torch}), whose inputs are tensors of data points, and the SMT solver (\texttt{f\_smt}), whose inputs are lists of symbolic variables. 
\end{sloppypar}

Following the model definition, we may outline the chosen certificate along with the relevant sets, as follows, where we assume to synthesise a quadratic control Lyapunov function over a spherical domain of radius 10.
\lstset{emph={%
    SYSTEM, DOMAINS, DATA, N_VARS, CERTIFICATE, TIME_DOMAIN, VERIFIER, ACTIVATION, N_HIDDEN_NEURONS, CTRLAYER, CTRLACTIVATION%
    },emphstyle={\color{redcode}\bfseries}%
}%

\begin{lstlisting}[language=Python, label=lst:test-file, caption=Example benchmark using Python-package interface.]
import fossil

# get the system model
open_loop = TestModel
system = fossil.control.GeneralClosedLoopModel.
            .prepare_from_open(open_loop())

# set the certificate domain
XD = fossil.domains.Sphere([0.0, 0.0], 10.)
sets = {fossil.XD: XD,}
data = {fossil.XD: 
        XD._generate_data(batch_size=500),}

# certificate and neural architectures parameters
opts = fossil.CegisConfig(
    SYSTEM=system,
    DOMAINS=sets,
    DATA=data,
    N_VARS=open_loop.n_vars,
    CERTIFICATE=fossil.CertificateType.LYAPUNOV,
    TIME_DOMAIN=fossil.TimeDomain.CONTINUOUS,
    VERIFIER=VerifierType.Z3,
    ACTIVATION=[fossil.ActivationType.SQUARE],
    N_HIDDEN_NEURONS=[4],
    CTRLAYER=[15, 1],
    CTRLACTIVATION=[fossil.ActivationType.LINEAR],
)

# start the synthesis process 
fossil.synthesise(opts)
\end{lstlisting}

\lstset{emph={%
    ahdjkk
    },emphstyle={\color{redcode}\bfseries}%
}%

The procedure first pre-processes the model (line 5) to include the dynamics within a closed-loop model. 
We then can define the domain set, a sphere centered at the origin (line 9). 
The domain set is used both in its symbolic formulation, for verification purposes, and as a set to sample datapoints from. 
These two distinct aspects are specified as \texttt{sets} including the symbolic set formulations, whilst \texttt{data} denotes the samples generated through the \texttt{\_generate\_data} method. 

Following the definition of the Lyapunov certificate and the time domain (lines 20-21), we can set a few additional parameters within the ad-hoc class \texttt{CegisConfig}.
We choose the Z3 solver as the SMT engine, 
the candidate certificate is embodied by a neural network with a single hidden layer of 4 neurons with square activation function. Note that by increasing the list of neurons, we increase the layers of the network: e.g. \texttt{[4, 5]} creates a network with two hidden layers composed of 4 and 5 neurons, respectively. 
Finally, we may specify the neural architecture of the control network, a single hidden layer of 15 neurons and 1 outputs (representing the single control input), with a linear activation (denoting a canonical feedback control law) -- naturally, the definition of a control architecture is not needed for autonomous models.

The command \texttt{synthesise} starts the procedure and its CEGIS loop. The default number of loops is set to 10, but can be easily modified by setting the additional parameter \texttt{CEGIS\_MAX\_ITERS} (not shown).
A detailed list of parameters (e.g., certificates, domain sets) supported by \fossil can be found in the parameters guide at the project's repository:  \href{https://github.com/oxford-oxcav/fossil}{https://github.com/oxford-oxcav/fossil} \cite{fossilrepo}.

\subsection{Extensibility of \fossil}
\label{subsec:use-extensibility}

\subsubsection{New Certificate-based Properties}
\fossil 2.0 is a tool for verifying properties of dynamical models using certificates. We provide a broad range of certificates for continuous-time models, but we appreciate that users may wish to synthesise certificates that prove properties not covered.  With this in mind, Fossil 2.0 presents a significantly improved codebase over Fossil 1.0 that enables extensions to new certificates. Here, we demonstrate how a new certificate can be specified within Fossil.

Let us first explain how Fossil's codebase is structured to enable defining further certificates. At its core, Fossil consists of sub-modules corresponding to the components described in Section~\ref{subsec:cegis}. 
The tasks of the \emph{learner} and \emph{verifier} must vary for each certificate: the learner must define a loss function that trains a neural network to satisfy the certificate's conditions while the verifier must falsify these conditions. 

\begin{figure}
    \centering
    \includegraphics{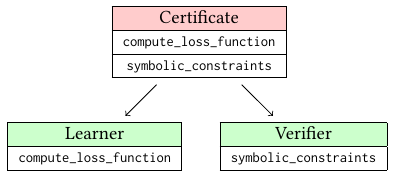}
    \caption{Schematic representation of the \texttt{Certificate} class providing required functionality to the components of CEGIS.}
    \label{fig:cert-module}
\end{figure}

We delegate the tasks specific to a given certificate to a new module, the \emph{certificate} module. A defined certificate must provide the following functionality: calculation of a loss function to guide learning; and the construction of the symbolic formula consisting of the negation of the conditions for the certificate to be valid. A schematic depiction of the certificate code structure is provided in Fig.~\ref{fig:cert-module}.

Recall the stability property defined in \eqref{eq:stab-spec}, and let us consider the Lyapunov certificate that proves this property for a continuous time dynamical model. 
Given a domain $\cX$ and a model $f: \cX \to \cX$ with unique equilibrium point $x^* \in \cX$, consider a function $V: \cX \subset \real^n \to \real, V \in \mathcal{C}^1$. $V$ is a Lyapunov function if:
\begin{subequations}
\label{eq:lyap}
\begin{align}
    &V(x^*) = 0,  \label{eq:lyap-v-zero-in-zero} \\
    &V(x) > 0 \quad \forall x \in \cX \setminus \{x^*\}, \\
    &\dot{V}(x) = \langle \nabla V(x), f(x) \rangle < 0 \quad \forall x \in \cX \setminus \{x^*\}.
\end{align}
\end{subequations}
We illustrate in Listing~\ref{lst:certificate} a class which defines the required functionality.  
The \texttt{compute\_loss} method calculates a dedicated loss function based on the conditions specified in Eq. \eqref{eq:lyap}, whilst the \texttt{get\_constraints} method returns the symbolic constraints relevant for the certificate (specifically the negation of the above conditions). 

\begin{lstlisting}[language=Python, caption={Pseudocode of a Certificate file.}, label=lst:certificate]
class LyapunovCertificate(Certificate):

def init(self, domain):
    # initialise the domain of verification
    self.domain = domains[XD]

def compute_loss(self, V, grad_V, f):
    """ Calculate loss function based on sample points
    - V: Values of certificate
    - grad_V: Values of gradient of certificate 
    - f: Values of vector field
    """
    lyap_loss = relu(-V).mean()
    Vdot = torch.sum(torch.mul(grad_V, f), dim=1)
    lie_loss = (relu(Vdot)).mean()
    loss = lyap_loss + lie_loss
    return loss

def get_constraints(self, verifier, C, Cdot):
    """ SMT-based constraints for Certificate conditions.
    - verifier: Verification object
    - C: Certificate formula
    - Cdot: Certificate lie derivative formula 
    """
    lyap_constr = _And(C <= 0, self.domain)
    lie_constr = _And(Cdot >= 0 self.domain)
    return lyap_constr, lie_constr
\end{lstlisting}

The loss function penalises positive values of $\dot{V}(x)$ and negative values of $V(x)$, hence the choice of the ReLU function. Other choices are possible: accordingly, our tool supports several loss function computations. 
The  \texttt{get\_constraints} method returns the negation of the symbolic conditions, as the verifier searches for an instance (a counter-example) that satisfies them. 

Notice that condition \eqref{eq:lyap-v-zero-in-zero} is not included in the certificate file: its satisfaction is automatically guaranteed by considering $x^*$ as the origin (the default setting), by choosing activation functions that evaluate to zero in $x^*$, and by omitting any network bias, thus ensuring $V(x^*) = 0$.

\subsubsection{Bespoke Domains}
A crucial limitation of the provided command line interface is that all domains specified must be one amongst a hyper-sphere, -torus or -box. Within the package, users may specify domains that are bespoke to their verification problem. This requires defining two methods: one which returns a symbolic expression representing the domain, and one which provides data points sampled over the domain. Examples of this may be found amongst the large number of benchmarks showcased at \cite{fossilrepo}.

\section{Experimental Evaluation}
\label{sec:experiments}
\fossil 2.0 greatly extends and fundamentally recasts the software library of Fossil 1.0, as presented in~\cite{abate2021FOSSILSoftwareTool}: this earlier work has compared Fossil 1.0 against competitive state-of-the-art techniques such as SOS-tools, and has shown to deliver a faster synthesis for the most commonly synthesised certificates for dynamical models: Lyapunov functions and barrier certificates. 

Hence, considering \fossil 1.0 as the state-of-the-art tool for formal synthesis of certificates, the presented experimental evaluation of \fossil 2.0 is twofold. 
Firstly, \fossil 2.0 is benchmarked against the capabilities of its predecessor in terms of computational efficiency \cite{edwards2023GeneralVerificationFramework}.
Secondly, we present an selection of benchmarks of Fossil 2.0 for certificates of the seven properties mentioned in Table \ref{tab:tool-features} \cite{edwards2023GeneralVerificationFramework}. 
We also present some new benchmarks for discrete-time models, in particular to prove their stability and safety.

\subsection{Comparison against \fossil 1.0}
\label{subsec:comparison-fossil-1}

We employ the benchmark suite originally outlined in \cite{abate2021FOSSILSoftwareTool}, which solely include Lyapunov and barrier functions. We use the same network structure (in terms of width and choice of activation function for each layer) for both tool versions to give a fair comparison. 

Each benchmark is repeated ten times, each run being initialised with a different random seeding. We consider two measures to determine the quality of each tool. Firstly, how often the tool correctly terminates successfully (having verified the property) out of 10 runs. We report this as the success rate $S$. Secondly, we report the average, minimum and maximum computation times over all successful runs. These results are collected in Table \ref{tab:fossil-comp}.

As shown in Table~\ref{tab:fossil-comp}, it is clear that the approaches are similar when synthesising the more straightforward Lyapunov function, with a slight  improvement in terms of speed and robustness.
Meanwhile for barrier certificate synthesis we achieve significant improvements in terms of both success rate and synthesis time relative to the baseline.
There are several factors accounting for this computation improvement, including an improved loss function, an enhanced communication between the CEGIS components, and an overall streamlined software implementation.

\begin{table*}[htbp]
    \centering
    \begin{tabular}{lllllrrrrrrrr}
    \toprule
              &       &             &          &                                                   & \multicolumn{4}{c}{Fossil 1.0} & \multicolumn{4}{c}{Fossil 2.0}                                                   \\
              &       &             &          &                                                   & $\min$                         & $\mu$                          & $\max$ & $S$    & $\min$ & $\mu$ & $\max$ & $S$ \\
    Benchmark & $N_s$ & Certificate & Neurons  & Activations                                       &                                &                                &        &        &        &       &        &     \\
    \midrule
    NonPoly0  & 2     & Stability   & [5]      & [$\varphi_{2}$]                                   & 0.04                           & 0.21                           & 1.58   & 100 & 0.01   & 0.16  & 1.47   & 100 \\
    Poly2     & 2     & Stability   & [5]      & [$\varphi_{2}$]                                   & 0.35                           & 11.71                          & 70.39  & 90  & 0.01   & 5.64  & 45.46  & 90  \\
    Barr1     & 2     & Safety      & [5]      & [$\sigma_{\mathrm{sig}}$]                         & 100.17                         & 100.17                         & 100.17 & 10  & 1.60   & 7.11  & 13.05  & 100 \\
    Barr3     & 2     & Safety      & [10, 10] & [$\sigma_{\mathrm{sig}}$,$\sigma_{\mathrm{sig}}$] & 16.80                          & 101.72                         & 334.79 & 50  & 13.42  & 29.35 & 84.55  & 100 \\
    \bottomrule
\end{tabular}

    \caption{Comparison of \fossil 1.0 vs \fossil 2.0 (the present work). Here, we use the same naming scheme for benchmarks as used in \fossil 1.0. $N_s$: Number of states. We show the \emph{Property} being verified and the network structure (in terms of \emph{Neurons} and \emph{Activations} in each layer). Symbol $\varphi_i$ denotes polynomial activations of order $i$, e.g. $\varphi_2$ identifies a square activation function; symbol $\sigma_{\textrm{sig}}$ denotes sigmoid activation.  We report success rate ($S$) and the minimum, mean ($\mu$) and maximum computation time $T$ over successful runs, in seconds.}
    \label{tab:fossil-comp}
\end{table*}

\subsection{Synthesis of Certificates}
\label{subsec:other-results}

\begin{table*}
    \centering
    \begin{tabular}{lrrllllllr}
    \toprule
       & $N_s$ & $N_u$ & Certificate & Neurons        & Activations                                                                      & \multicolumn{3}{c}{$T$ [s]} & Successful (\%)                        \\
       &       &       &             &                &                                                                                  & $\min$                      & $\mu$           & $\max$         & $S$ \\
    \midrule
    1  & 3     & 0     & Stability   & [10],          & [$\varphi_{2}$],                                                                 & 0.18 ($\approx$0.0)                  & 0.45 ($\approx$0.0)      & 1.23 (0.02)    & 100 \\
    2  & 2     & 2     & Stability   & [5],           & [$\varphi_{2}$],                                                                 & 0.09 (0.01)                 & 0.25 (0.02)     & 0.53 (0.03)    & 100 \\
    3  & 3     & 3     & ROA         & [8],           & [$\varphi_{2}$],                                                                 & 1.31 (0.02)                 & 42.27 (0.03)    & 312.12 (0.04)  & 100 \\
    4  & 8     & 0     & Safety      & [10],          & [$\varphi_{1}$],                                                                 & 31.36 (7.48)                & 124.29 (31.64)  & 172.47 (43.15) & 70  \\
    5  & 3     & 1     & Safety      & [15],          & [$\sigma_{\mathrm{t}}$],                                                         & 1.65 (0.18)                 & 12.31 (2.37)    & 53.94 (7.1)    & 90  \\
    6  & 3     & 0     & SWA         & [6], [5]       & [$\varphi_{2}$], [$\sigma_{\mathrm{t}}$]                                         & 0.2 (0.04)                  & 2.66 (0.09)     & 13.12 (0.18)   & 90  \\
    7  & 2     & 1     & SWA         & [8], [5]       & [$\varphi_{2}$], [$\varphi_{2}$]                                                 & 0.06 (0.03)                 & 0.19 (0.09)     & 0.58 (0.22)    & 90  \\
    8  & 3     & 0     & RWA         & [16],          & [$\varphi_{2}$],                                                                 & 1.41 (0.1)                  & 15.18 (0.13)    & 78.57 (0.19)   & 90  \\
    9  & 2     & 1     & RWA         & [4, 4],        & [$\sigma_{\mathrm{sig}}$,$\varphi_{2}$],                                         & 0.57 (0.24)                 & 6.58 (3.03)     & 19.54 (9.98)   & 100 \\
    10 & 3     & 0     & RSWA        & [16],          & [$\varphi_{2}$],                                                                 & 5.09 (0.12)                 & 28.82 (0.18)    & 86.5 (0.25)    & 100 \\
    11 & 2     & 2     & RSWA        & [5, 5],        & [$\sigma_{\mathrm{sig}}$,$\varphi_{2}$],                                         & 1.0 (0.15)                  & 1.26 (0.25)     & 1.66 (0.42)    & 100 \\
    12 & 2     & 0     & RAR         & [6], [6]       & [$\sigma_{\mathrm{soft}}$], [$\varphi_{2}$]                                      & 6.81 (1.02)                 & 25.34 (6.16)    & 79.55 (14.41)  & 100 \\
    13 & 2     & 2     & RAR         & [6, 6], [6, 6] & [$\sigma_{\mathrm{sig}}$,$\varphi_{2}$], [$\sigma_{\mathrm{sig}}$,$\varphi_{2}$] & 5.45 (1.31)                 & 27.55 (9.44)    & 106.25 (57.1)  & 100 \\
    \midrule
    14 & 2     & 0     & Stability   & [2],           & [$\varphi_{2}$],                                                                 & 0.05 ($\approx$0.0)                  & 0.51 (0.02)     & 1.85 (0.04)    & 100 \\
    15 & 2     & 2     & Stability   & [5],           & [$\varphi_{2}$],                                                                 & 0.04 ($\approx$0.0)                  & 2.66 (0.02)     & 16.11 (0.04)   & 100 \\
    16 & 2     & 0     & Safety      & [2],           & [$\varphi_{2}$],                                                                 & 0.36 (0.32)                 & 0.62 (0.54)     & 1.26 (0.81)    & 100 \\
    17 & 2     & 2     & Safety      & [10, 10],      & [$\sigma_{\mathrm{sig}}$,$\varphi_{2}$],                                         & 0.17 (0.04)                 & 0.21 (0.1)      & 0.28 (0.16)    & 100 \\
    \bottomrule
\end{tabular}

    \caption{Results of synthesis on our benchmark portfolio. Here, columns take the same meaning as in Table \ref{tab:fossil-comp}. In addition, $N_u$: number of inputs; $\sigma_{\mathrm{soft}}$: softplus function, $\sigma_{\mathrm{t}}$: hyperbolic tangent. We index benchmarks by number in the first row. The \emph{SWA} and \emph{RAR} properties are proven by certificates of two functions; comma-separated lists show the structures for each function.
    }
    \label{tab:benchmarks-results}
\end{table*}

We test our tool over 13 continuous-time benchmarks and 4 discrete-time models, covering each of the properties mentioned in Section \ref{subsec:properties}. 
Since the initialisation of the network is random and it may affect the performance of the overall procedure, 
for each benchmark tests are repeated 10 times, with different initial random seeds for each repeat. 
Recall that our technique is not guaranteed to terminate: we thus set a maximum of 
100 cegis loops for SWA and RAR properties, and 25  CEGIS loops for the other certificates; if the procedure reaches this limit, the run is counted as a failure. 
We highlight that our tool is able to perform well on easier (linear, polynomial) and nontrivial (in presence of trigonometric and transcendental functions) benchmarks in terms of computational time and consistency of successful synthesis. 
Due to the large number of benchmarks, and their relative complexity, details on the benchmarks may be found in a benchmarks document at the corresponding artifact, see \cite{fossilrepo}. 

Table~\ref{tab:benchmarks-results} shows the results on our benchmarks portfolio. We select one autonomous and one control model per property, with $N_s$ and $N_u$ denoting the number of state variables and the number of control inputs, respectively. 
We report the number of neurons and activation functions, where symbol $\varphi_i$ denotes polynomial activations of order $i$ -- e.g.  $\varphi_2$ indicates a square activation function -- and $\sigma_{sig}$, $\sigma_{soft}$, $\sigma_{t}$ indicates sigmoid, softplus, tanh functions, respectively. For brevity, the control architecture is omitted from the table; we briefly describe them next.
All control networks are structured with one hidden layer, composed of less than 10 neurons for all case studies; we employ both linear and nonlinear (hyperbolic tangent) activation functions, prompting linear and nonliner control designs. 
We report the success rate under the $S$ column and the average time ($\mu$, over the 10 runs), the minimum and maximum computational times (in seconds) under the column $T$. In brackets, we also denote the amount of time spent during the learning phase of our procedure, with approximately all remaining time spent during the verification phase.

We highlight that our tool synthesises the certificate for the several supported properties with high success rate (consistently close to 100\%) within a few seconds, witnessing the robustness (and the practical termination) of our proposed method. 
These results also showcase the flexibility of the approach, which handles polynomial certificates, closer to traditional control applications, and non-polynomial certificates, more in line with machine learning frameworks.  

\subsection{A Case Study}
In order to further demonstrate the usage of \fossil 2.0, we present next a case study for a complex certificate. Consider the following (controllable) dynamical model of an inverted pendulum with friction and two control inputs:
\begin{equation} 
\label{eq:inv-pend}
    \begin{cases}
        \dot{x}_0 = x_1 + u_0, \\
        \dot{x}_1 = u_1 + (mL^2)^{-1} (mgL \sin(x_0) - b \cdot x_1),
    \end{cases}
\end{equation}
where $m$ and $L$ are the respective mass and length, and $g$ and $b$ are physical constants representing gravity and friction. We aim at proving a reach-avoid remain property for this model, which qualitatively means proving that all trajectories starting in some initial set reach some goal set, while remaining within some safe region (or correspondingly, avoid an unsafe region). Once reaching the goal set, trajectories must then remain within a final set for all time. With this case study, we are challenging \fossil 2.0 to synthesise two neural-based certificates concurrently with feedback control laws, in order to prove that the closed loop model satisfies the desired specification.

We shall demonstrate construction the configuration file. 
First, let $g=9.81$, $b=0.1$, $m=0.15$ and $L=0.5$. We can declare the dynamics
\begin{lstlisting}[language=yaml, numbers=none]
SYSTEM: [x1 + u0, u1 + (0.73575*sin(x0) - 0.1*x1) / (0.0375)]
\end{lstlisting}
Next, we must define within the \fossil input file the reach-avoid-remain problem we wish to solve. Here, we consider a relatively simple problem of trajectories starting within some box around the origin, reaching and then remaining within smaller boxes all while staying within some larger safe box. We characterise the problem here using a safe set, as opposed to the unsafe set in Eq.~\eqref{eq:rar-spec}, though this is dually equivalent. 
\begin{lstlisting}[language=yaml, numbers=none]
CERTIFICATE: RAR
DOMAINS:
  XD: Rectangle([-3.5, -3.5], [3.5, 3.5])
  XS: Rectangle([-3.0, -3.0], [3.0, 3.0])
  XI: Rectangle([-2.0, -2.0], [2.0, 2.0])
  XG: Rectangle([-0.1, -0.1], [0.1, 0.1])
  XF: Rectangle([-0.2, -0.2], [0.2, 0.2])
TIME_DOMAIN: CONTINUOUS
VERIFIER: DREAL
\end{lstlisting}

\begin{figure}
    \centering
    \includegraphics[width=\linewidth]{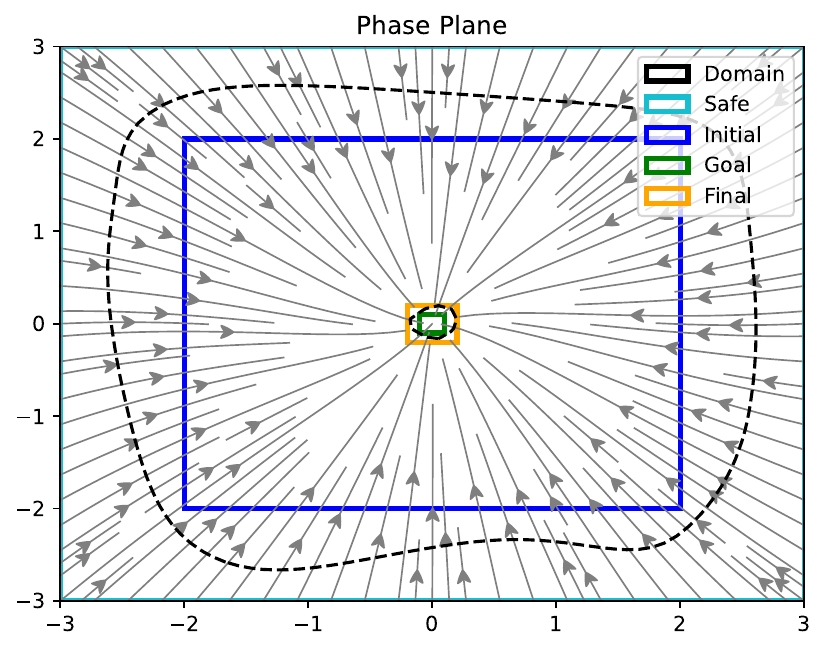}
    \caption{Phase plane of the closed loop model for the presented case study, as well as the zero contours of the two functions that comprise the reach-avoid-remain certificate.}
    \label{fig:rar_casestudy}
\end{figure}

Finally, we declare the configuration for learning. Notably, a reach-avoid-remain certificate consists of two different functions (i.e. two different networks are being trained and consequently  verified), which are synthesised concurrently as a proof. We specify the structure for the second function composing the certificate using the \texttt{N\_HIDDEN\_NEURONS\_ALT} and \texttt{ACTIVATION\_ALT} fields. 
In addition, we shall also specify the desired feedback control structure: we consider here a simple linear feedback control. 
\begin{lstlisting}[language=yaml, numbers=none]
N_HIDDEN_NEURONS: [6]
ACTIVATION: [SIGMOID, SQUARE]
N_HIDDEN_NEURONS_ALT: [6]
ACTIVATION_ALT: [SIGMOID, SQUARE]
CTRLAYER: [8,2]
CTRLACTIVATION: [LINEAR]
\end{lstlisting}
Finally, we can pass this configuration to \fossil, which takes 50s to successfully synthesise a controller and certificate. We depict the obtained certificate (via the zero contours of its two constituent functions) and  trajectories of the closed-loop model in the phase plane diagram shown in Fig. \ref{fig:rar_casestudy}.

\section{Discussion and Conclusions}
\label{sec:concl}

We have presented \fossil 2.0, a software tool for the verification of properties of dynamical models via automated formal synthesis of a broad range of certificates, based on recent advancements in the field of certificate synthesis. Certificate synthesis is based on a CEGIS loop, exploiting neural networks to provide candidate functions, which are then formally verified with the help of SMT solvers.

\fossil 2.0 greatly expands on the feature-set of \fossil 1.0, incorporating a much broader selection of certificate-based verification queries for dynamical models. Furthermore, \fossil 2.0 is able to concurrently synthesise controllers which guide a model to satisfy a specification in parallel to a certificate that proves the property holds.
\fossil 2.0 is  further endowed with an enhanced pythonic interface over its predecessor, as well as a easy-to-use command line interface for casual users.

We present results for a number of benchmarks showcasing its diverse specification portfolio and robustness to initialisation, and results comparing the novel release to the state-of-the-art tool \fossil 1.0, which it is able to outperform consistently. 

In future extensions of \textsf{Fossil}, we hope to further explore the modular synthesis of certificates, and to improve its extensibility to additional model semantics, in particular models that are stochastic.
Further, future efforts will be directed towards improved scalability by e.g. additional means of verification, and the implementation of an automated choice of activation functions.

\begin{acks}
    The authors would like to thank Daniele Ahmed and Dr. Mirco Giacobbe for their contributions to the development of the original release of \fossil \cite{abate2021FOSSILSoftwareTool}. Alec was supported by the EPSRC Centre for Doctoral Training in Autonomous Intelligent Machines and Systems (EP/S024050/1).
\end{acks}

\bibliographystyle{ACM-Reference-Format}
\bibliography{main}

\end{document}